# Water-based and Inkjet Printable Inks made by Electrochemically Exfoliated Graphene

Khaled Parvez[1,*], Robyn Worsley[1], Adriana Alieva[1], Alexandre Felten[2], Cinzia Casiraghi[1,*]

[1] School of Chemistry, University of Manchester, Oxford Road, Manchester M13 9PL, UK

[2] University of Namur, Rue de Bruxelles 61, 5000 Namur, Belgium

**Abstract**

Inkjet printable graphene inks are very attractive for applications in flexible and foldable electronics, such as wearable electronics and the Internet of Things. However, the ink preparation is still very time consuming as high concentrations can be achieved only with prolonged sonication (>24 hours) or with expensive setups. Here we demonstrate a water-based inkjet printable ink made from electrochemically exfoliated graphene. A printable and stable (> 1 month) ink with concentration of ~2.25 mg mL$^{-1}$ was formulated in less than 5 hrs, using two successive steps: first exfoliation and dispersion of large graphene flakes (> 5 μm) followed by 1 hour tip-sonication to reduce the flake size below 1 μm, as required for the material to be ejected by the nozzle. The formulated ink contains more than 75% single- and few-layers (i.e. less than 2 nm in thickness) graphene flakes with an average lateral size of 740 nm. Thermal annealing allows to achieve high C/O ratio (>10), which translates into one of the highest electrical conductivity (≈3.91 × 10$^4$ S m$^{-1}$) reported so far for solution-processed graphene, without the use of any harsh chemical processing.

**Corresponding Authors:** E-mails: khaled.parvez@manchester.ac.uk (Khaled Parvez); cinzia.casiraghi@manchester.ac.uk (Cinzia Casiraghi)

## 1. Introduction

A new generation of lightweight, flexible and foldable electronics has emerged for applications in areas such as wearable electronics and the Internet of Things. In parallel, the field of printed electronics has developed rapidly, driven by printing techniques offering low-cost and simple methods for devices fabrication, whilst demonstrating compatibility with most substrates, including those that are soft and flexible [1]. In particular, inkjet printing is an additive, non-contact printing technique that allows for excellent scalability, low wastage and design flexibility [2]. This technique has been successfully used to deposit a wide range of nanomaterials [1, 3-5].

Graphene, a carbon sheet with atomic thickness, is an attractive material for printed electronics due to its high conductivity, thermal and chemical stability and flexibility [6]. Typically, inkjet printable graphene inks are prepared using liquid-phase exfoliation (LPE) [7], where delamination of graphite is directly promoted via ultrasound waves [7] or high-shear forces [8, 9] in a medium that can stabilize graphene flakes (e.g. organic solvents, aqueous surfactant solutions, etc.) without introducing covalent functionalization. In particular, LPE is a very attractive technique to make inkjet printable formulations [10-15] of graphene because it provides pristine graphene (i.e. oxygen-free) with size well below that of the nozzle, by minimizing nozzle blockage. However, LPE still shows the following limits: in the case of bath sonication, the process is very time consuming (typically, several days) and the exfoliation efficiency strongly depends on the ultrasonic transducers design: the ultrasound intensity is not uniformly distributed in the medium, badly affecting the batch-to-batch reproducibility [16]. Tip sonication allows reducing the exfoliation below 24 hrs [16] – however, the direct positioning of the tip in the medium results in strong material fragmentation (i.e. breaking of bonds in the nanosheets) [16], which is deleterious for use of materials in electronics, where large and thin flakes are preferred to small and thick ones due to lower flake-to-flake junction resistance obtained for larger flakes and due to the unique properties of single-layer graphene. Thus, bath sonication is usually preferred to tip-sonication for graphene exfoliation [16]. Tip sonication has also shown to decompose solvents such as N-methyl pyrrolidone (NMP) [17]. Sonochemical degradation can potentially affect the stability of solvents and stabilizers, by lowering the efficiency and reproducibility of the obtained formulations. Finally, tip-sonication may result in higher costs as the sonic probe wears down on prolonged sonication and a chiller should be preferred to an ice bath, which needs to be replenished every 2 hrs or so [16]. Nevertheless, the concentration of graphene dispersion produced via LPE method in various organic solvent is typically below 0.5 mg mL$^{-1}$ [18]. Although a concentration up to 1.2 mg mL$^{-1}$ has been reported, this was achieved with extremely long sonication time (i.e. 460 hrs) [19]. Additionally, exfoliation with shear mixer typically gives low concentrations of few-layers graphene [8], which are not ideal to make an inkjet printable ink. Alternative mass scalable approaches to LPE have been recently reported, such as ball-milling [20, 21] and micro-fluidization [22, 23]. The yield of thin graphene layers obtained by ball-milling is low (< 1wt%) [20], and the scalability and batch-to-batch reproducibility of this method have not been proved yet. Moreover, the concentration of graphene dispersions obtained by ball-milling typically do not exceed 0.5 mg mL$^{-1}$ [21]. In case of micro-fluidization, the setup is relatively expensive, the processing time is determined by the piston passes (i.e. 2 to 4 hrs) and the exfoliated flakes show structural defects, whose concentration increases with the piston passes [22]. It is therefore important to develop alternative exfoliation techniques to develop inkjet printable inks, possibly making use of low cost and non-toxic solvents such as water. However, making an inkjet printable formulation in water is a challenge on its own as water does not have the correct rheological properties to be used as printable solvent [24]. A recent work from our group reported a water-based solvent that allows producing inkjet printable inks of graphene and analogous material [24]. In this protocol, the material is still produced by LPE, making the in preparation rather time consuming (e.g. exfoliation time > 5 days).

Electrochemical exfoliation (ECE) [25] is a simple and scalable technique, which involves the use of an electrolyte and an electrical current to encourage structural delamination of a graphite electrode. In particular, anodic ECE allows achieving exfoliation in few minutes [26] and to achieve more than 70% single- and bi-layers with large lateral size (1-10 μm) [27]. Despite numerous reports on graphene formulations in water [28-31] prepared by electrochemical exfoliation, there are only a few works demonstrating inkjet printable formulations of graphene produced by ECE. Miao et al. [32] and Li et al. [33] reported inkjet printing of electrochemically exfoliated graphene (ECG) by dispersing the material in dimethylformamide (DMF) and cyclohexanone based solvents, respectively. Although printability in these solvents was achieved, the high boiling point, toxicity and flammability of these solvents limit their widespread use in printed technologies.

Here we demonstrate a water-based inkjet printable ink made from electrochemically exfoliated graphene (ECG). The printable ink production is achieved in less than 5 hrs, leading to a stable (for over a month) formulation with concentration of 2.25 mg mL$^{-1}$. The formulated ink contains more than 75% single and few-layers (i.e. less than 2 nm in thickness) graphene flakes with an average lateral size of 740 nm. Such ink formulation allows stable jetting, rapid ink drying (less than 10 s) as well as ensuring wetting of untreated substrates such as paper and glass with negligible printing failure. Thermal annealing allows to achieve high C/O ratio (>10), which translates into one of the highest electrical conductivity ($\approx 3.91 \times 10^4$ S m$^{-1}$) reported, without the use of any harsh chemical processing.

## 2. Experimental section

**2.1. *Electrochemical exfoliation of graphene***: The ECG was produced according to the procedure described in our previous work [27]. Briefly, two pieces of graphite foil (Alfa Aesar, 0.13 mm thick) were inserted into 0.5 M $(NH_4)_2SO_4$ aqueous electrolyte solution as cathode and anode at a distance of ≈2 cm, respectively. The electrodes were then biased at a DC voltage of 10 V for the exfoliation. The voltage was kept constant for 5 min. During this process, the graphene flakes were readily exfoliated from the anode. The exfoliated material was collected through a cellulose nitrate filter membrane (pore size 0.2 μm, Whatman) and rinsed several times with de-ionized (DI) water by vacuum filtration. Finally the material was dispersed in DI water by bath sonication for 30 min.

**2.2. *Graphene ink formulation***: To formulate the ECG ink, we first reduce the flake size of the ECG by using a tip-sonicator (Fisher Scientific, FB50) for 1 h. The use of such a short tip-sonication time allows avoiding problems related to fragmentation (i.e. defects formation) and possible sonication degradation. Diameter of the probe was 1/8 inch and the ultrasonic frequency was 20 kHz. In order to keep constant temperature, the vessel was placed in an ice bath during sonication. After sonication, the ECG dispersion in water was centrifuged using a Sigma 1-14k refrigerated centrifuge at 903 g for 20 min to remove residual graphite particles. Finally, to make ECG dispersion suitable for inkjet printing, the rheology of water was changed by adding surface tension and viscosity modifiers and a binder, as reported in ref [24], using mild stirring for 10 min. The change of solvent rheology can be done using a solvent exchange process (i.e. exchanging the solvent from water to the previously-prepared water-based printable solvent) or by introducing the additives directly in water, leading to a total ink preparation time between 3 and 5 hrs, depending on the protocol used.

For comparison, pristine graphene ink were also prepared via liquid-phase exfoliation (LPE) method, where 1.5 g of graphite flakes (Sigma Aldrich, 100+ mesh) and 0.5 g of 1-Pyrenesulfonic acid sodium salt (PS1, also from Sigma Aldrich) were mixed into 500 mL de-ionized (DI) water [34]. The mixture was then sonicated at 600 W using a Hilsonic bath sonicatior for 5 days. Afterwards, the dispersion was centrifuged with two different steps to remove un-exfoliated graphite and excess PS1 [24].

**2.3. *Characterization of ECG ink***: UV-Vis spectroscopy was used to determine the final concentration of the ECG ink. Measurements were performed using a Perkin-Elmer I-900 UV-Vis-NIR spectrometer (without integrating sphere) using an optical glass cuvette having a cell length of 1 cm. In case of LPE based graphene, the concentration was extracted using the Beer Lambert law and an extinction coefficient of 2460 $L\ g^{-1}\ m^{-1}$ measured at 660 nm [7]. Note that the precise extinction coefficient value is still matter of discussion and new values have been recently reported [35]. As the size and chemical structure of ECG is different from graphene produced by LPE, it may be inaccurate to use the same extinction coefficient. Thus, we have directly measured the extinction coefficient of ECG by diluting the dispersion and by measuring the weight of material. We found an extinction coefficient of 2207 $L\ g^{-1}\ m^{-1}$ which is in good agreement with that of LPE-based graphene. Thus, the size and relatively small amount of oxygen in the material does not seem to strongly affect the extinction coefficient.

A Bruker Atomic Force Microscope (MultiMode 8) in Peak Force Tapping mode, equipped with ScanAsyst-Air tips is used to determine the lateral size and thickness distribution of the flakes. The sample was prepared by drop casting the solution on a clean silicon substrate; several areas of 1600 µm$^2$ were scanned and about 201 flakes were selected for lateral size analysis. The same sample preparation has been used for Raman measurements. About 20 isolated flakes were measured. Raman measurements were performed using a Renishaw Invia Raman spectrometer equipped with a 514.5 nm excitation line with 1 mW laser power. 100X NA0.85 objective lens and 2400 grooves/mm grating were used for measurements. The Raman spectra were fitted with Lorentzian line shapes and the background was subtracted by using a linear fit under each peak. The X-ray photoelectron spectroscopy (XPS) measurements were performed on an Escalab 250Xi spectrometer from Thermo Scientific. A monochromatic K-alpha source (1486.6 eV) was used. The spectra were acquired using a spot size of 400 microns and constant pass energy (150 eV for survey and 20 eV for high resolution spectra). A flood gun with combined electrons and low energy Ar ions is used during the measurements.

**2.4.** *Inkjet printing of ECG ink*: A Dimatix DMP-2800 inkjet printer (Fujifilm Dimatix, Inc., Santa Clara, USA) was used. The printer can create and define patterns over an area of about 200 mm × 300 mm and can handle substrates up to 25 mm thick. The nozzle plate consists of a single raw of 16 nozzles of 21 µm diameter spaced 254 µm apart with a typical drop volume of 10 pL. We used several substrates: PEL P60 paper (from Printed Electronics Ltd.), characterized by a micro-porous surface treatment, designed to wick away the carrier solvent of the ink, while allowing for uniform deposition. In addition, clean microscope glass slides (1.2 mm thick) were also used. Prior to inkjet printing, the glass slides were cleaned with both acetone and 2-propanol for 15 min each in an ultrasonic bath, followed by drying with N$_2$. The ECG inks are printed at a voltage of ~30 V using a jetting frequency of 5 kHz and with drop spacing of 25 µm and 35 µm on paper and glass substrates, respectively. The pulse length during the inkjet printing was 35.6 µs. An array of ECG ink drops (1 mm × 1 mm) were also printed on Si/SiO$_2$ substrate (Siegert

Wafer GmbH) with 80 μm drop spacing. During the inkjet printing of ECG ink on paper, the substrate temperature was kept at room temperature (i.e. ≈25 °C), while printing on glass was carried out with a substrate temperature of 40 °C, to facilitate the rapid drying of the ink.

**2.5. *Electrical characterization***: All the current-voltage (I-V) characteristics of the fully printed ECG ink on both paper and glass substrates were taken with a two contact probes using an Agilent B1500 probe station. The sheet resistance ($R_s$) values were calculated from the I-V profiles using the equation; $R_s = R \times W/L$, where, R, W and L are the resistance, width and length of the printed patterns, respectively. The DC conductivity was calculated after measuring the thickness of the printed patterns by using a Bruker DektakXT surface profiler.

## 3. Results and discussion

Thin graphene flakes were produced by ECE [27], as schematically illustrated in **Figure 1a**. The exfoliated product was collected by vacuum filtration and repeatedly washed with water to remove any residual electrolyte. The collected powder was then dispersed in DI water without use of additional stabilizer/surfactants by sonication for 30 min (Supporting information **Figure S2a**). The flakes produced by ECE have lateral size larger than 1 μm (**Figure S2b**), in agreement with previous reports [27]. However, an inkjet printable inks requires the particles lateral size to be less than 1/50 of the nozzle diameter in order to avoid nozzle blockage [36]. Here we use a nozzle diameter of ≈21 μm, which gives an optimum flake size ≈400 nm. Therefore, in order to reduce the size of the flakes, we further sonicated the prepared ECG dispersion in water by tip-sonication for one hour (**Figure S4**). The use of such a short time tip-sonication allows reducing the flake size, without causing structural damage of the material and reproducibility issues caused by the tip's wear. The final step consists in changing the rheology of water, to make it suitable for inkjet printing [24] (see Experimental section for details).

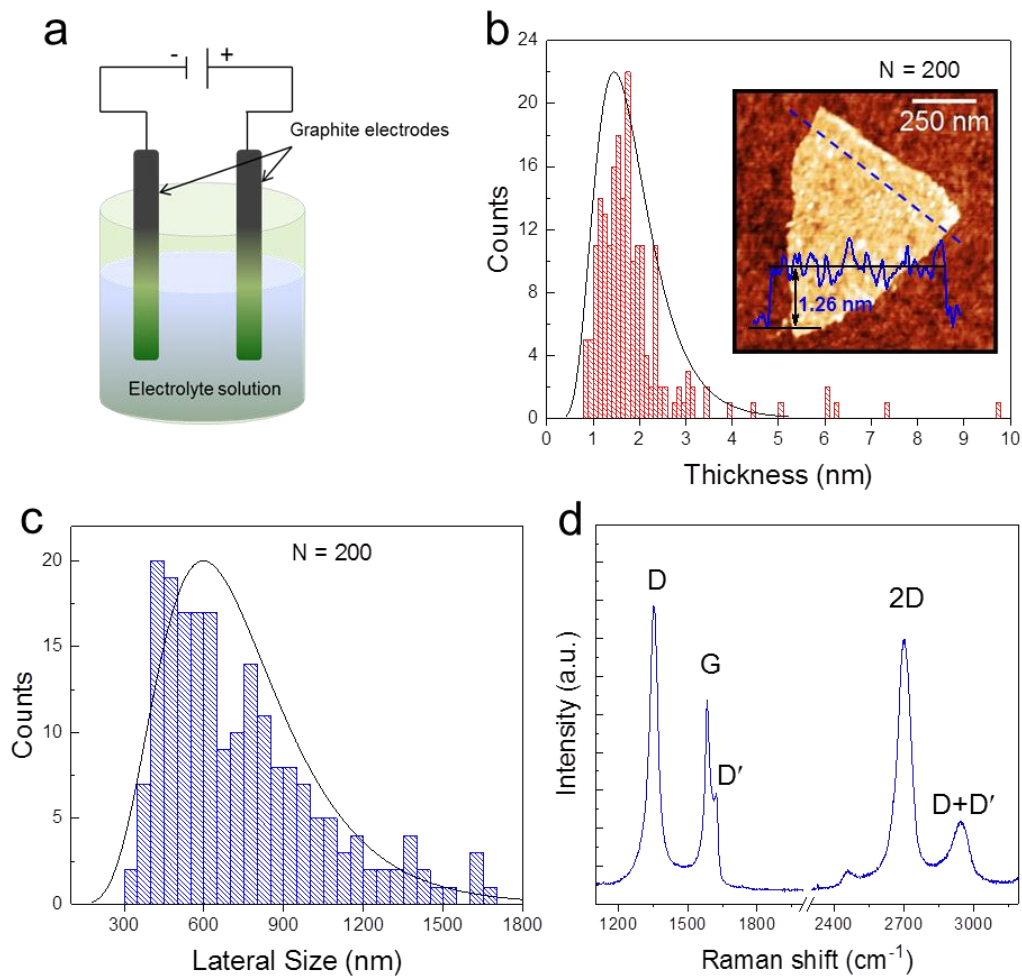

**Figure 1**: (a) Schematic illustration of the ECE setup. Statistics derived from AFM image of ECG flakes on Si/SiO$_2$, indicating: (b) thickness and (c) lateral flake size distribution derived from 200 ECG flakes, respectively. Inset of (b) shows the corresponding AFM image of a single ECG flake and corresponding profile showing a mean thickness of ~1.3 nm. (d) Representative Raman spectrum of ECG measured on an isolated flake.

The lateral size ($\langle L \rangle$) and thickness ($\langle t \rangle$) distributions of the ECG graphene flakes was estimated by atomic force microscopy (AFM). **Figure 1b** shows the statistics of peak thickness extracted from AFM over 200 individual flakes of graphene (examples in Supplementary **Figure**

**S2**). The log-normal distribution is peaked at a thickness of 1.86 nm. Considering the C-O functional groups on the surface due to ions intercalation, which is known to strongly oxidize graphene [29, 37], we can assume ~0.8 nm thickness for a single layer [25], thus the average number of layers is ≈2, in agreement with previous works [27, 38]. The ⟨L⟩ distributions of the flakes were also investigated to verify that the flake dimensions match with the inkjet printer requirements. The lateral size of the nanosheets follows a broad distribution from 300 nm to ~1.6 μm with a log-normal distribution peaked at ≈740 nm (**Figure 1c**). Although the ECG flakes are slightly larger than the 1/50 of nozzle diameter (i.e. ≈400 nm), previous reports [14, 24] have shown that graphene flakes with 1/20 (≈1 μm) of nozzle size can be successfully inkjet printed. This is probably related to the dimensionality and flexibility of graphene, allowing for easier ejection from the nozzles, compared to traditional spherical nanoparticles.

Raman spectroscopy is a powerful technique for the characterization of graphene [39]. **Figure 1d** shows a representative Raman spectrum measured on an individual ECG flake. The typical Raman spectrum shows the typical D and G peaks at ~1350 cm$^{-1}$ and ~1580 cm$^{-1}$, respectively, which have been observed in the Raman spectrum of ECG [27]. A relatively sharp 2D peak is located at ~2697 cm$^{-1}$ with full width at half-maximum (FWHM) of 66 cm$^{-1}$. In addition, a D′ peak is observed at 1622 cm$^{-1}$. The D peak is activated by defects [40-43]. In particular, the Raman spectrum of defective graphene can be described with a phenomenological three-stage amorphization trajectory [41]. In stage 1, starting from pristine graphene, the Raman spectrum evolves as follows: the D peak appears and the intensity ratio between D and G peaks, $I_D/I_G$ increases; the D' appears; all the peaks broaden and G and D' begin to overlap [41]. In this stage, $I_D/I_G$ can be used to estimate the amount of defects [41], while the intensity ratio between D and D' peaks ($I_D/I_{D'}$) can be used to distinguish between different type of defects [42]. At the end of Stage 1, the G and D' peaks are no more distinguishable, $I_D/I_G$ starts decreasing. As the number of defects keeps increasing, the Raman spectrum enters Stage 2, showing a marked decrease in the G peak position and increase broadening of the peaks; $I_D/I_G$ sharply decreases towards zero and second order peaks are no longer well defined [41]. Stage 3 describes amorphous materials with increasing $sp^3$ content [44]. The broad 2D peak (FWHM ~66 cm$^{-1}$) and high $I_D/I_G$ (~1.9, measured at 514.5 nm excitation wavelength) observed in Figure 1d demonstrate that ECE graphene is a defective graphene belonging to stage 2 [42]. Note that in case of highly defected graphene, the 2D peak FWHM cannot be used to identify single layers, as defects cause broadening of the peak, as electron-defect scattering lowers the life time of excited charge pair. The defects activating the D peak are likely to be the C-O functional groups introduced during intercalation as the flakes have average size still larger than the laser spot (~400 nm), thus contribution to the D peak from the edges should be negligible. In order to verify that the short sonication time does not damage graphene, i.e. no structural defects are introduced, we measured the Raman spectrum of ECG obtained with and/or without the second step of tip-sonication. **Figure S3** shows that the Raman spectra of the two ECG have comparable $I_D/I_G$ (~1.9) suggesting that the tip-sonication step, used to decrease the size of the flakes, does not introduce defects (at least within the Raman resolution). Note that, the flake size is >500 nm in both cases, so effects from the edges are negligible. We further confirmed this result by comparing the graphene obtained with and without tip-sonication by XPS. **Figure S6** shows the deconvoluted C1s spectra of ECG before and after tip-sonication. The carbon and oxygen content before tip-sonication is 80.9 % and 16.6%, respectively. After 1 h tip-sonication both the carbon and oxygen remained unchanged (i.e. 81.6% and 16.1%, respectively), thus demonstrating that the tip-sonication step does not introduce any additional defect on ECG, but it only reduces the flake size.

Optical absorption spectroscopy was used to estimate the concentration of ECG ink. In order to determine the concentration of the ECG ink, the extinction coefficient ($\varepsilon$) was first determined experimentally (see Experimental section for details). Relationship between absorption per unit length and known graphene concentrations is shown in **Figure S7a**. The slope of the straight line fit through the data points in **Figure S7b** provided the extinction coefficient value of $\varepsilon = 2207$ L g$^{-1}$ m$^{-1}$. Using the measured extinction coefficient and the Beer-Lambert law [45], we estimated a concentration of ~2.25 mg mL$^{-1}$,(**Figure S8a**) which is high enough to print conductive graphene patterns with few printing passes [24]. Moreover, the formulated ECG ink remains stable without any significant precipitation under ambient condition for more than a month (**Figure S8b**).

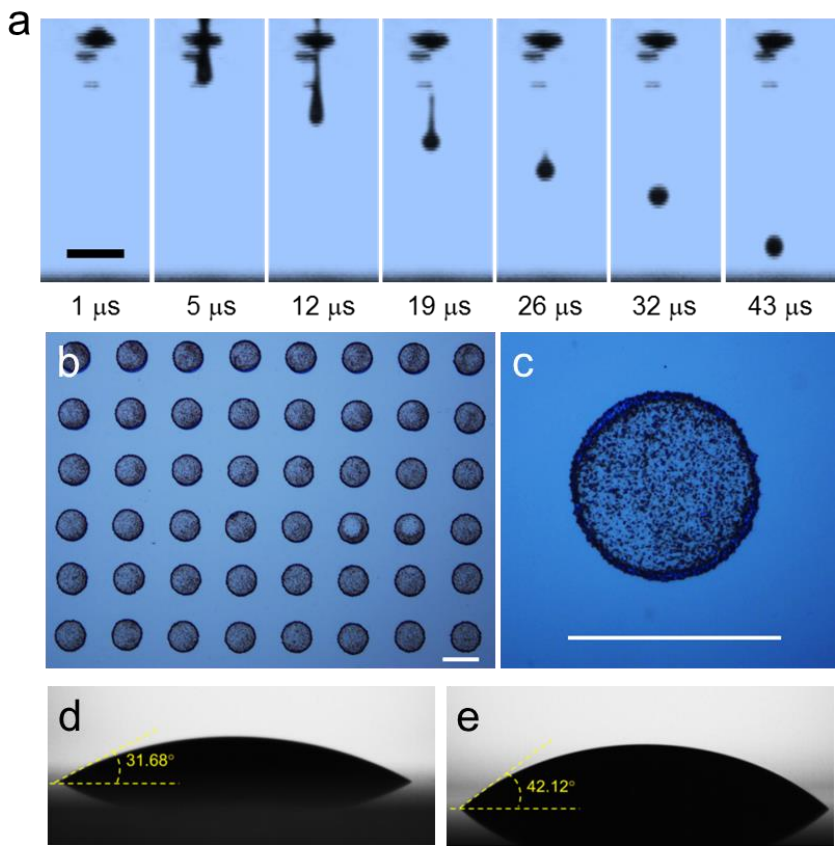

**Figure 2**: (a) Droplet jetting sequence observed from the printer camera as a function of time. Scale bar is 100 μm. (b) An array of printed graphene dots of ≈50 μm in size printed on Si/SiO$_2$ substrate. (c) Magnified image of a single printed dot; Scale bars in both (b) and (c) are 50 μm. (d) and (e) Contact angle of the ECG graphene ink on glass and Si/SiO$_2$ substrate, respectively.

Stable jetting is dependent on the ink viscosity ($\eta$), surface tension ($\gamma$) and density ($\rho$) as well as nozzle diameter ($\alpha$) of the cartridge, typically combined into an inverse Ohnesorge (Oh) number, $Z = Oh^{-1} = (\gamma\rho\alpha)^{1/2}/\eta$ [46]. The ink is expected to produce stable drops if $1 < Z < 14$, with $Z < 1$ indicate an ink that will not jet and $Z > 14$ an ink prone to generation of secondary droplets [46]. We formulated the ECG ink with the following rheological parameters: $\eta_{ECG}$ 1.72 mPa.s, $\gamma_{ECG}$ ≈51 mN m$^{-1}$ and $\rho_{ECG}$ ≈1.0 g cm$^{-3}$. These values give $Z \approx 19$ for our water based ECG ink. Despite $Z > 14$, the drop is stable and no satellite drops or nozzle blocking were observed, as confirmed by the jetting sequence of the ECG ink (**Figure 2a**). This is in agreement with our previous report on water-based inks of 2D materials, which show good printability with $Z \approx 20$ [24]. **Figure 2b** shows the printed drops matrix on Si/SiO$_2$ with a resolution (i.e. droplet diameter) of ≈50 μm. The as-printed droplets exhibit good uniformity and are free from misplaced drops, confirming the printing reliability. **Figure 2d and e** show the wetting behaviour of the ECG ink, where the average contact angles were found to be ≈32° and ≈43° on glass and on Si/SiO$_2$ substrates, respectively.

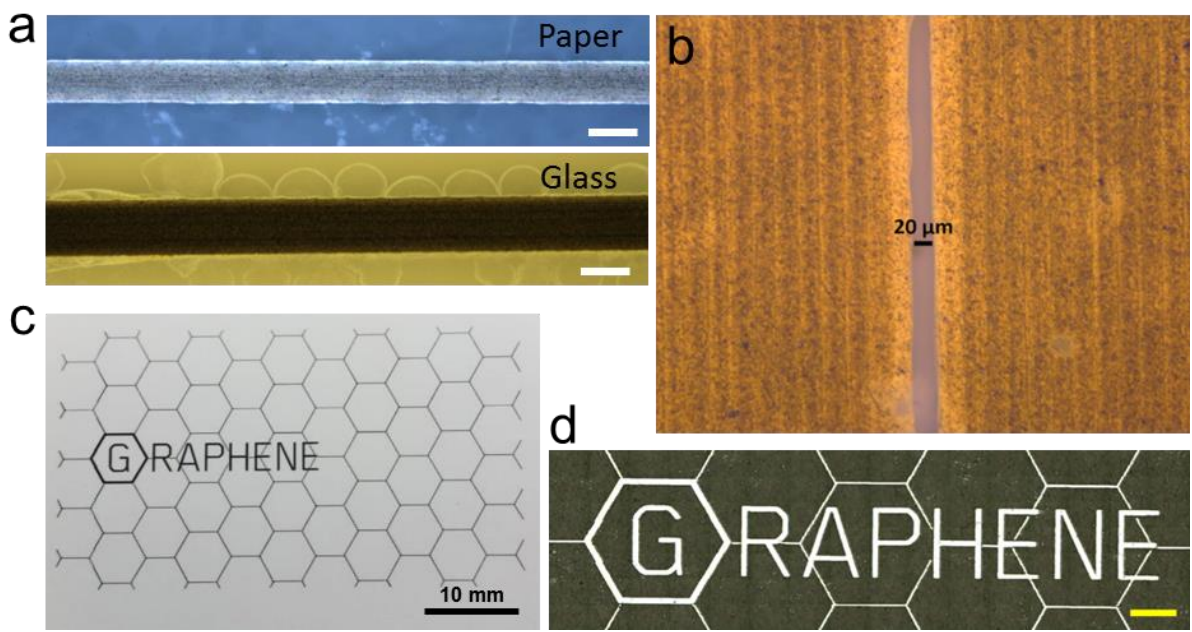

**Figure 3**: (**a**) Optical micrographs of printed tracks on paper and glass substrates. Scale bar 200 μm. (**b**) Two graphene contacts printed with a gap of ≈20 μm on PEL P60 paper substrate. (**c**) Inkjet printed 'Graphene' logo on paper substrate over an area of 50 mm x 30 mm and, (**d**) Zoom of word 'graphene' in (**c**), showing high precision and uniform printing of our ink; scale bar is 2 mm.

**Figure 3a** shows optical micrographs of the graphene lines printed on both paper and glass substrates. The graphene flakes are evenly deposited along the length and width of the lines without any noticeable coffee rings. No substrate heating was used during the printing on paper, while the glass substrate was heated at 40 °C during the printing processes, since these printing conditions not only allow optimal morphology, but also rapid ink drying. Indeed we observed drying of ink in less than 10 s from the printed fiducial camera. Moreover, we were able to achieve high resolution printing where two graphene contacts were printed with a gap of only ≈20 μm on paper (**Figure 3b**). We then investigated printing of ECG graphene on large scale. **Figure 3c and d** shows a uniform, highly resolved inkjet printed graphene logo on paper over a 50 mm × 30 mm area by using only 3 print passes. This clearly demonstrates that our ECG ink can be used to fabricate printed devices on large area without the loss of printing resolution, nozzle blockage and with excellent material uniformity.

To investigate the electrical performance of the printed patterns, centimeter size graphene lines at various numbers of passes were printed on both paper and glass substrates with a fixed concentration of 2.25 mg mL$^{-1}$. As shown in **Figure S9a**, the printed films became darker as the number of printing passes increased from 5 to 50. Highly uniform and continuous patterns were obtained after 10 print passes (**Figure S9b**).

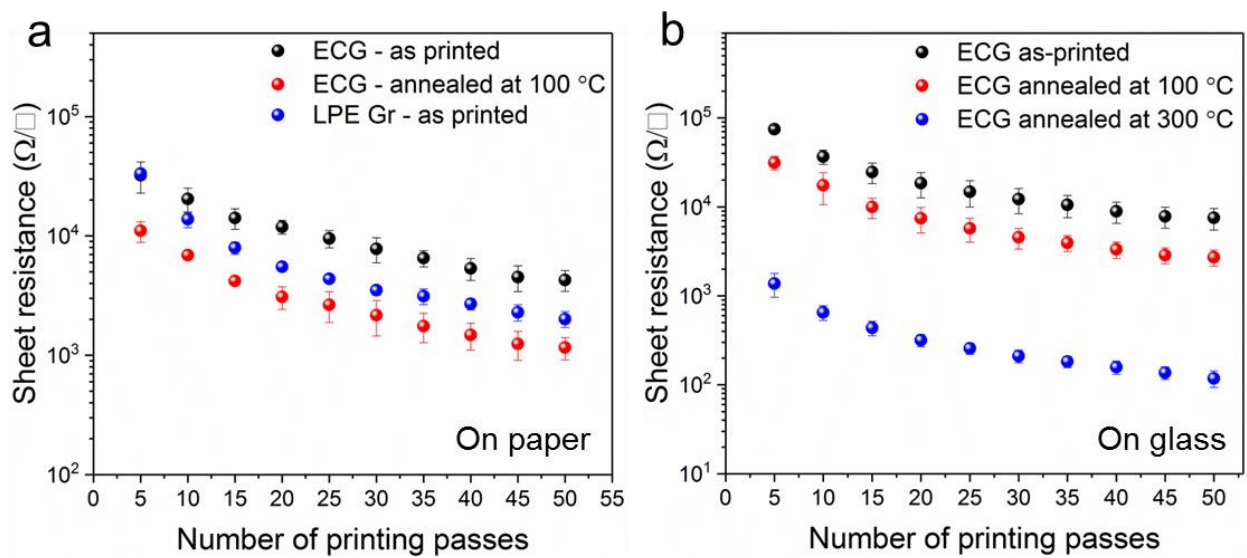

**Figure 4**: Variation of sheet resistance with number of printing repetition (**a**) on paper and (**b**) on glass substrate. Ink concentration: ~ 2.25 mg mL$^{-1}$.

Two-terminal current-voltage (I-V) measurements show that all the printed graphene lines exhibited linear Ohmic characteristics (**Figure S10 and S11**), where the current increases with increasing number of printing passes. **Figure 4b and c** show the values of sheet resistance ($R_s$) of the printed graphene lines on paper and glass, respectively, for an increasing number of printed passes. After the percolation threshold is reached [47], $R_s$ decreases rapidly until the percolation to bulk threshold transition, as observed in our previous work [24].

Let us compare the $R_s$ of the ECG graphene with that of the graphene ink produced by LPE both printed under the same conditions (i.e. ink concentration ~2.23 mg mL$^{-1}$(**Figure S1**), drop spacing 25 µm and no-platen heating) on paper. **Figure 4a** shows that after 50 print passes, the $R_s$ of ECG reaches only ≈4.3 ± 0.8 kΩ □$^{-1}$ without any post-treatment and is comparable with LPE pristine graphene ink (i.e. ≈2.0 ± 0.3 kΩ □$^{-1}$). After annealing at 100 °C for 1 h in vacuum, the $R_s$ of the ECG films is ≈1.2 ± 0.2 kΩ □$^{-1}$, smaller than that of graphene produced by LPE. As for the graphene lines printed on glass substrate, the $R_s$ reaches ≈2.7 ± 0.6 kΩ □$^{-1}$ using 50 print passes and annealing at 100 °C (**Figure 4b**). The slightly higher $R_s$ of the graphene lines on glass substrate compared to the printed lines on paper are mainly due to the larger drop spacing used for printing on glass (i.e. 35 µm) compared to the drop spacing used on paper (i.e. 25 µm). The technical paper (i.e. PEL P60) used in this work is relatively smooth and porous, therefore it allows to deposit the material on a smaller area, giving rise to lower sheet resistance, as observed in our previous work [24]. Nevertheless, after annealing the printed graphene at 300 °C for 1 h under continuous $N_2$ flow, the I-V profile shows Ohmic behaviour with much higher current (**Figure S11c**). The $R_s$ value of all the graphene patterns significantly decreases with increasing number of printing passes (**Figure 4b**): the lowest $R_s$ obtained is only ≈118 Ω □$^{-1}$ using 50 printing passes. The line profile measured across the printed ECG with different printing passes (**Figure S12a**), revealed reasonably uniform thicknesses with sharp edges along the cross-sections. Moreover, the mean height of the printed lines plotted as function of number of passes shows that the thickness of the printed graphene pattern increases linearly with increasing number of printing passes, where the average layer thickness for each printing passes is found to be ≈4.5 nm on glass substrate (**Figure S12b**). The RMS roughness of the ECG after 50 print passes is ~85 nm (**Figure S12c**) which is slightly higher than the previously reported LPE graphene based ink [48]. **Figure S13** shows the DC conductivity (σ) of the inkjet printed ECG graphene patterns on glass substrates. After 50 printing passes, the average thickness is ≈223 nm, leading to a conductivity of $3.91 \times 10^4$ S m$^{-1}$. Note that the electrical performance of our printed graphene film is well superior to that in other works reporting ECG inks. For example in ref. [32], for a film thickness of ~250 nm the sheet resistance reported is above $10^5$ Ω □$^{-1}$, while in our case the sheet resistance is below $10^4$ Ω □$^{-1}$ (**Figure 4b**).

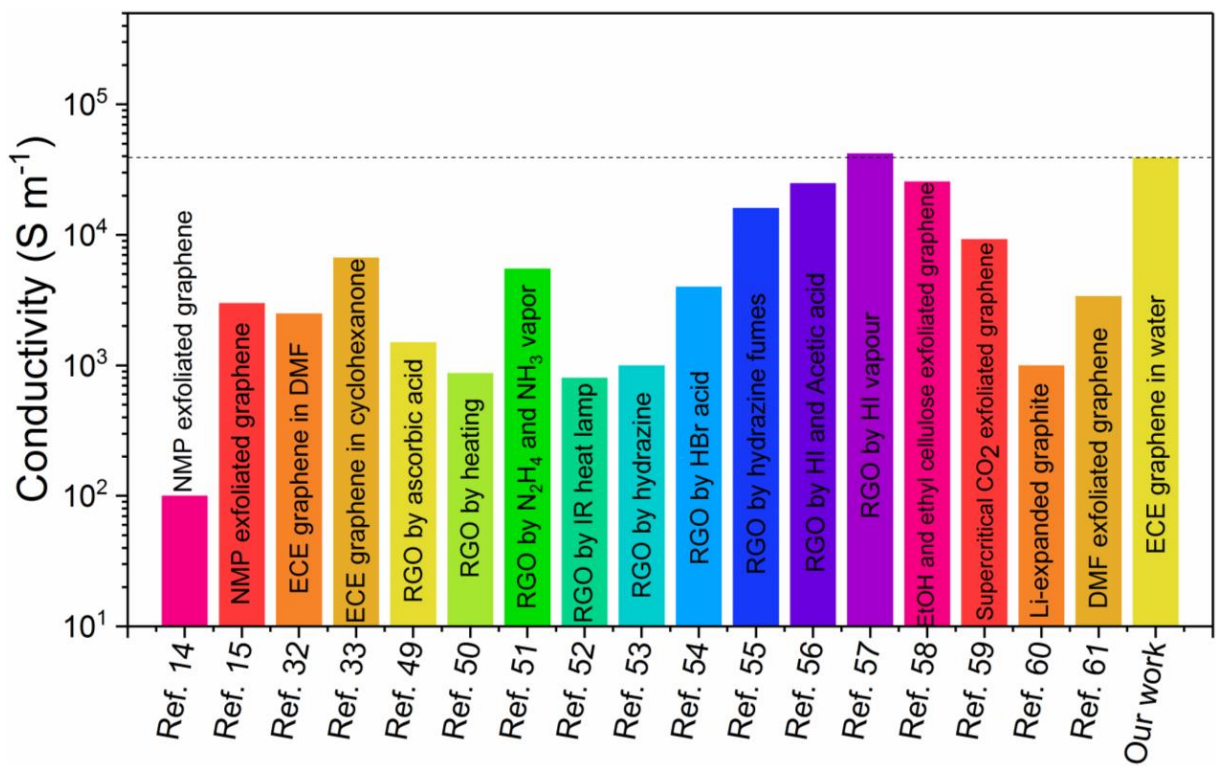

**Figure 5**: Comparison of the electrical conductivity measured on our inkjet printed electrochemically exfoliated graphene (ECG), after annealing at 300 °C for 1 h, with literature results based on different types of inkjet printable graphene formulations.

**Figure 5** and **Table S1** compare the electrical conductivity of our ink, after annealing at 300 °C for 1 h with those reported for other types of inkjet printable formulation of oxygen-free graphene. We also include for reference reduced graphene oxide (rGO) based printable inks.[49-57] The maximum conductivity reported for printed pristine graphene inks made by LPE is

≈2.56 × 10$^4$ S m$^{-1}$ [58], while most reported values are < 1.0 × 10$^4$ S m$^{-1}$ [14, 15, 59-61]. The highest conductivity of inkjet printed graphene is based on HI vapour rGO, which is ≈4.2 × 10$^4$ S m$^{-1}$, resulting from a film thickness above 1 μm [57], well higher than the typical thickness reported in our work (i.e. ~223 nm). For oxygen-free graphene, the highest value of conductivity is obtained in ref. [52], where inks are prepared by dispersion of graphene/ethyl cellulose powder in a solvent system composed of 85:15 v/v cyclohexanone/terpineol. Note that the use of relatively high amount of ethyl cellulose is likely to disrupt printing (e.g. nozzle blockage, separation of the binder from graphene after solvent evaporation, etc.) and the solvents used are toxic. Moreover, the reported electrical conductivity values of inkjet printable ECE based graphene inks are less than 1.0 × 10$^4$ S m$^{-1}$ [32, 33], where the highest conductivity value is ~6.7 × 10$^3$ S m$^{-1}$ [33]. Therefore, the electrical conductivity achieved in this work (i.e. 3.91 × 10$^4$ S m$^{-1}$) is higher than previously reported ECE based graphene inks, as well as most of printable inks made with LPE and rGO.

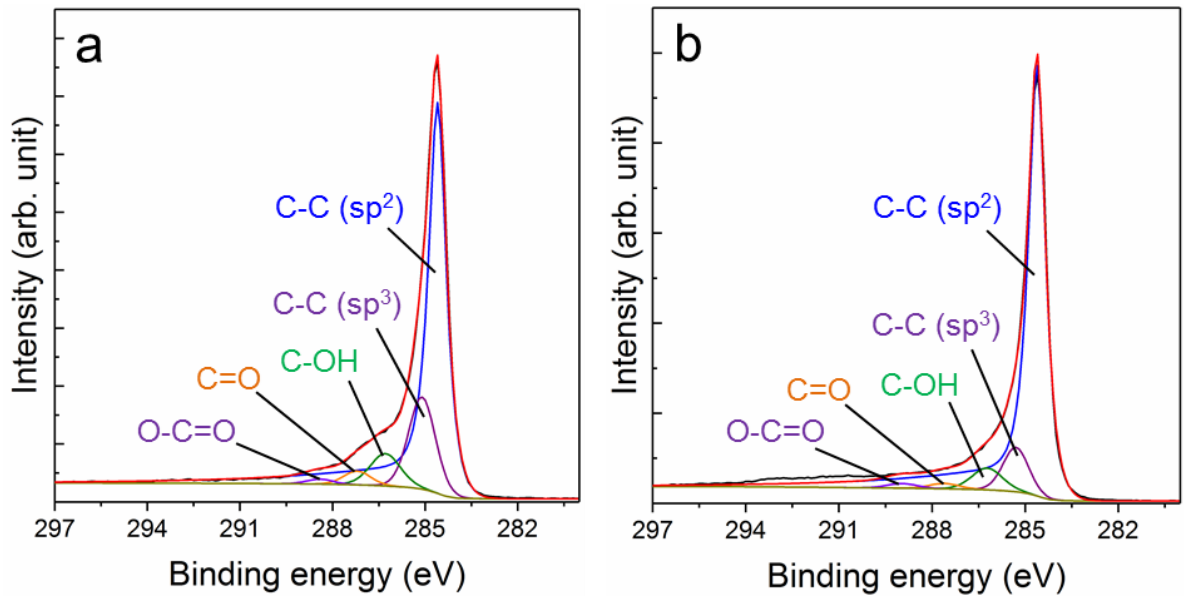

**Figure 6**: XPS spectra of ECG, after annealing at (a) 100 °C and (b) 300 °C for 1 h, respectively.

In order to understand the stability of the ECG graphene in printable water-based solvent as well as the high electrical performance of the printed patterns, structural and chemical composition of the film after annealing were measured by XPS. **Figure 6a, b** shows the deconvoluted C1s spectra of the annealed ECG at 100 and 300 °C, respectively. Deconvoluted spectra shows five different characteristics peaks at 284.4 ± 0.2, 285.1 ± 0.2, 286.5 ± 0.3, 287.4 ± 0.2 eV and 288.8 ± 0.4 eV corresponding to C-C ($sp^2$), C-C ($sp^3$), C-OH, C=O and O-C=O functional groups [62]. The C=C peak refers to the $sp^2$ carbon components, while oxygen containing functional groups correspond to the amount of $sp^3$ hybridized carbon. XPS measurements show that as-prepared ECG contains approximately 16 atom% of oxygen functional groups, yielding a C/O ratio of 5.07 (**Figure S6b**). Note that GO typically has oxygen content above 30% [63]. The C/O ratio of our ECG is higher than that of typical GO produced by means of Hummers method (C/O ratio ~2), but is slightly lower than that of reduced GO (C/O ratio >7) [64]. Annealing of ECG at 100 °C, slightly improves the C/O ratio (~8.3), whereas the amount of oxygen containing functional groups are significantly reduced after annealing the ECG at 300 °C for 1 h (**Figure 6c**), revealing a C/O ratio of ~10.4. Therefore, the low oxygen content of ECG after annealing results in high electrical conductivity of the inkjet printed ECG, comparable to rGO (**Figure 5**). In addition to the low oxygen content, the relatively large flake size (i.e. ~740 nm) of the ECG ink compared to the flake size of LPE and rGO based inks (i.e. < 500 nm) reported in the literature (**Figure S14** and **Table S1**), can result in lower flake-to-flake junction resistance, yielding to higher electrical conductivity. Note that there are some reports on inkjet printing of rGO with flakes > 1 μm [55-57]; however, the flakes contain a high defect densities due to harsh chemical treatment used for reduction and hence lower conductivity compared to that of graphene produced by LPE and ECE. Furthermore, one has to take into account that the different chemical composition of the inks, e.g. the type and amount of co-solvent or additives used to make the formulation, can affect printability and also electrical conductivity, as shown in **Figure S14**, where a clear correlation between size of the flakes and conductivity is not observed. In addition, the morphology of the inkjet printed graphene network also plays a crucial role. This can change depending on the deposition method. For example, in our previous work, we showed that the inkjet printed films have more oriented and ordered structure that can give rise to better electrical properties, compared to films made by vacuum filtration [48]. Because of that, **Table S1** and **Figure S14** only report works based on ink-jet printing.

To sum up, our exfoliation protocol offers a faster, safer and environmentally friendly production of inkjet printable graphene inks to be exploited for several applications in printable electronics.

## 4. Conclusion

In summary, we demonstrated a water-based and inkjet printable ink made from electrochemically exfoliated graphene (ECG). The printable ink production is achieved in less than 5 hrs, leading to a stable (for over a month) formulation with concentration of 2.25 mg mL$^{-1}$. The formulated ink contains more than 75% single and few-layers (i.e. flakes with less than 2 nm in thickness) graphene with an average lateral size of 740 nm. Such ink formulation allows stable jetting, rapid ink drying (less than 10 s) as well as ensuring wetting of untreated substrates such as paper and glass with negligible printing failure. The sheet resistance is better than that obtained by water-based inks made by LPE, using the same concentration and printing parameters, after annealing. Furthermore, thermal annealing at 300 °C allows to achieve high

C/O ratio (>10), which translates into one of the highest electrical conductivity ($\approx 3.91 \times 10^4$ S m$^{-1}$) compared to the reported inkjet printable LPE, rGO and ECE based graphene inks.

**Notes**

The authors declare no competitive financial interest.

**Acknowledgment**

CC and KP acknowledge the Grand Challenge EPSRC grant EP/N010345/1. RW acknowledges the Hewlett-Packard Company for financial support in the framework of the Graphene NowNano Doctoral Training Centre. AA acknowledges financial support by the European Research Council (ERC) under the European Union's Horizon 2020 research and innovation programme (grant agreement No 648417).

**Appendix A. Supplementary data**.

Supplementary data to this article can be found online at
https://doi.org/10.1016/j.carbon.2019.04.047.